\documentclass[aps,prl,reprint,twocolumn,superscriptaddress,floatfix]{revtex4-2}

\usepackage{amsmath} % include math package
\usepackage{amssymb} 
\usepackage{wasysym}
\usepackage{graphicx}% Include figure files
\usepackage{dcolumn}% Align table columns on decimal point
\usepackage{bm}% bold math
\usepackage[hidelinks]{hyperref}
\usepackage{color} % for colored text
\usepackage{siunitx}
\usepackage{comment}
\usepackage[T1]{fontenc} % for Michals name

\newcommand{\fig}[1]{Fig.~\ref{fig:#1}}
\newcommand{\Fig}[1]{Fig.~\ref{fig:#1}}
\newcommand{\Figs}[1]{Figs.~\ref{fig:#1}}
\newcommand{\Eq}[1]{Eq.~\eqref{eq:#1}}
\newcommand{\Supp}[1]{SM}
\newcommand{\Var}{\mathrm{Var}}

\begin{document}

\title{Chromato-axial memory effect in step index multimode fibers}

\author{Louisiane Devaud}
\affiliation{Laboratoire Kastler Brossel, ENS-Université PSL, CNRS, Sorbonne Université, Collège de France, 24 rue Lhomond, 75005 Paris, France}

\author{Marc Guillon}
\affiliation{Saints-Pères Paris Institute for the Neurosciences, CNRS UMR 8003, Université de Paris, 45 rue des Saints-Pères, Paris, 75006, France}

\author{Ivan Gusachenko}
\affiliation{Laboratoire Kastler Brossel, ENS-Université PSL, CNRS, Sorbonne Université, Collège de France, 24 rue Lhomond, 75005 Paris, France}

\author{Sylvain Gigan}
\affiliation{Laboratoire Kastler Brossel, ENS-Université PSL, CNRS, Sorbonne Université, Collège de France, 24 rue Lhomond, 75005 Paris, France}
%\author{to put?}
%\affiliation{Institut Universitaire de France (IUF), Paris, France}

\date{\today}

\begin{abstract}
Multimode fibers (MMF) are used in many applications from telecomunications to minimally invasive micro-endoscopic imaging.
However, the numerous modes and their coupling make light-beam control and imaging a delicate task.
To circumvent this difficulty, recent methods exploit priors about the transmission of the system, such as the so-called optical memory effect.
Here, we quantitatively characterize a chromato-axial memory effect in step-index MMF, characterized through its slope $\delta z/\delta \lambda$ and its spectral and axial widths.
We propose a theoretical model and numerical simulations in good agreement with experimental observations.

\end{abstract}

\maketitle

Multimode fibers (MMF) are ubiquitous tools having a key role in telecommunications~\cite{richardson2013space}, and driving research in many other related fields such as MMF lasers~\cite{wright2017spatiotemporal}, fiber-based tunable optical cavities~\cite{steinmetz2006stable} and reconfigurable linear operators in quantum photonics applications~\cite{leedumrongwatthanakun2020programmable,matthes2019optical}. MMF have also been shown to be useful as high-resolution spectrometers~\cite{redding2014high,redding2013all}.
In addition, MMF have recently raised major hopes for developing non-invasive imaging techniques and minimally invasive endoscopes~\cite{choi2012scanner,papadopoulos2013high}.
Due to the complex nature of multi-mode guided propagation, light field control through MMF has highly benefited from recent progresses in wavefront shaping techniques, initially developed for adaptive optics in astronomy~\cite{roddier1999adaptive,tyson2015principles} and then extended to complex media~\cite{popoff2010measuring}.
As for propagation through complex media, uncontrolled coherent light propagation through a MMF results in a speckle intensity pattern~\cite{papadopoulos2012focusing,goodman_fundamental_1976}.
Light-field control at the distal tip of the fiber demands iterative optimisation~\cite{di2011hologram} or transmission matrix (TM) measurement~\cite{popoff2010measuring,vcivzmar2012exploiting}.
Not only this calibration is experimentally delicate, but it is also very sensitive to perturbations such as mechanical or thermal fluctuations.
In this regard, significant progresses have been made for controlling short MMF eigenmodes, resulting in particular in an increased robustness to bending~\cite{ploschner2015seeing,matthes2020learning}.
New strategies have also been developed to help TM measurement, using continuous optimisation~\cite{caravaca2013real} or exploiting priors~\cite{li2020compressively}.

A major prior information about the TM of complex media is the so-called "optical memory effect" (ME), referring to optical-field transforms which commute with the TM, or equivalently, which are diagonal in the eigen-basis of the TM~\cite{li2020compressively}.
The expression ME has been coined in the context of scattering media where for a thin enough diffuser, wavefront tilting has been demonstrated, both theoretically~\cite{feng_correlations_1988} and experimentally~\cite{freund_memory_1988}, to be preserved along propagation.
This spatial ME has then been extended to scattering media with strong anisotropy factors~\cite{judkewitz_translation_2015,osnabrugge_generalized_2017}.
Based on the cylindrical symmetry of MMF, rotational ME could be demonstrated~\cite{amitonova_rotational_2015,li2021memory}.
Recently, the concept of ME has even been broadened to the spectral domain~\cite{vesga2019focusing,zhu2020chromato}.
It has been shown experimentally that a chromatic shift induces an axial drift of a beam focused by wavefront shaping behind $1~{\rm mm}$-thick brain tissues, over spectral widths as large as $\sim 100~{\rm nm}$~\cite{vesga2019focusing}.
Theoretical modeling established that this broadband chromato-axial ($\chi$-axial) ME is a characteristic of forward scattering media thinner than the transport mean free path~\cite{zhu2020chromato}, where the product $\lambda z$ (wavelength, axial plane), is specifically conserved when illuminated by a plane wave~\cite{zhu2020chromato,Arjmand2021three}.
In essence, the $\chi$-axial ME is due to the conservation of the transverse component of the wave-vector under spectral detuning.
Interestingly, the transverse wave-vector conservation naturally holds in step-index (SI) MMF~\cite{vcivzmar2012exploiting}.
Furthermore, broadband light propagation through a MMF has recently been observed resulting in an axially extended focus allowing efficient volumetric imaging~\cite{pikalek2019wavelength,turcotte2020volumetric}.
In this letter we experimentally characterize the $\chi$-axial ME, so far unexplored, at the distal facets of SI-MMFs, and theoretically model it based on the study of the correlation function.
The two eigen-axes of the correlation function, diagonal wherein the TM of the MMFs, are analytically derived, together with their respective widths.

 \medskip

\begin{figure}[t]
\centering
    \includegraphics[page = 4,width= 1\columnwidth]{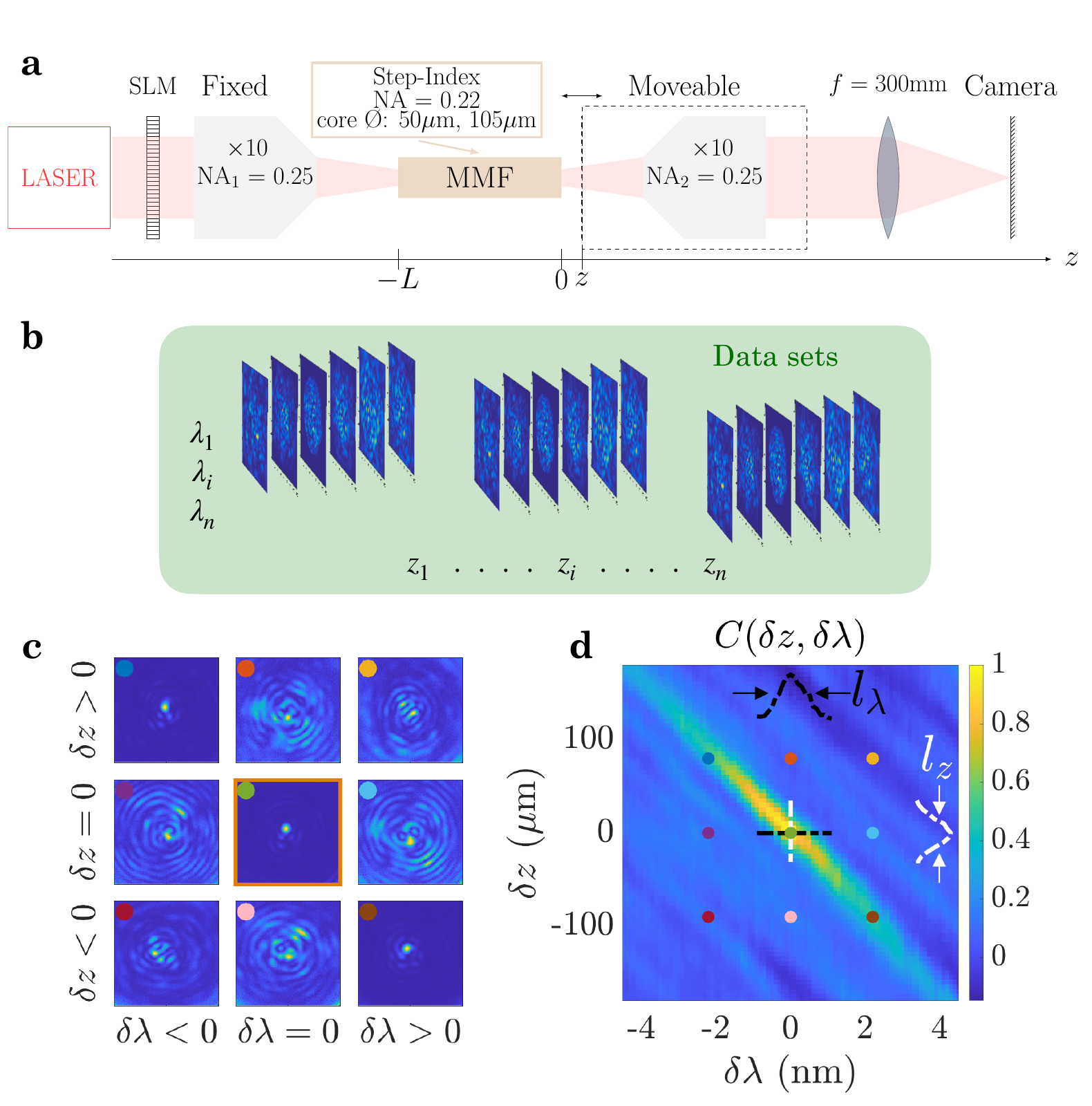}
    \caption{Experimental setup and procedure
    (a) A monochromatic tunable laser illuminates a MMF.
    The output filed at different axial positions is images on a CCD camera.
    A reflective phase-only SLM shapes the input wavefront (here represented as a transmission one for schematic convenience).
    (b) For each wavelength a microscope objective axial scan is performed and a set of images is acquired.
    (c) Observation of $\chi$-axial effect.
    A TM is measured at $\lambda_0$ = \SI{800}{nm} and $z_0 = 0$ and a focusing phase mask is applied on the SLM (central panel).
    The focus remains varying both the wavelength and the axial position.
    (d) Correlation plot of speckles acquired for the same fiber as in (c).
    A high-valued antidiagonal correlation region is visible.
    The colour dots make the correspondence with the images of (c).
    The widths of the profiles along the z axis ($l_z$) and along the $\lambda$ axis ($l_{\lambda}$) are presented on the side.
    }
    \label{fig:experimental_setup}
\end{figure}

As illustrated in~\fig{experimental_setup}a our experiment consists in a tunable mode-locked Ti:Sapphire laser emitting a linearly polarized monochromatic light coupled to a SI-MMF with a microscope objective.
In all experiments, MMFs of 0.22 numerical aperture (NA) and of lengths L~=~\SI{29}{\milli \meter}, \SI{58}{\milli \meter} and \SI{120}{\milli \meter} are tightly hold straight.
The objectives' NA ($0.25$) are chosen slightly larger than the fibers ones to ensure that light is well coupled to all fiber modes.
At the MMF output an objective lens mounted on a axial-translation stage collects the light and sends it to a CCD camera through a tube lens.
This arrangement enables imaging different axial planes at the MMF output, while tuning the wavelength in the range $800\pm 5~{\rm nm}$, as illustrated in~\fig{experimental_setup}b.
The longitudinal axis is denoted as $z$ and the position $z = 0$ is defined as the MMF output facet.
The input wavefront can be controlled by a spatial light modulator (SLM) conjugated with the back focal plane of the input objective in order to either refocus the beam after the MMF or generate random input fields.

In a first set of experiments, we engineered a focused \SI{800}{nm} laser spot just after the MMF by input wavefront modulation, using the TM of the fiber.
The TM is measured by sequentially displaying a basis of modes onto the SLM and gradually phase stepping them~\cite{popoff2010measuring}.
The focus is then generated by phase conjugation.
When applying a spectral shift to the input beam, the focus is observed to be axially shifted at the MMF output (see \fig{experimental_setup}c), similarly to what was observed through forward scattering media~\cite{vesga2019focusing}.
Noteworthy, over the explored spectral range, the contribution of the dispersion of the SLM and other optics to this effect is only a few percent~\cite{vesga2019focusing} as discussed in the Supplemental Materials~\cite{SM}.
The $\chi$-axial ME is not only observed for an engineered focused spot, but also when coupling random light patterns to the fiber.
In this case, the similarity of speckles can be observed by computing the zero-mean cross-correlation products between output speckles~\cite{zhu2020chromato}:
\begin{equation}
    C(\delta z,\delta \lambda) =  \frac{\langle \tilde{I}(\delta z,\delta \lambda) \tilde{I_0} \rangle}{\sqrt{\langle \tilde{I}^2(\delta z, \delta \lambda) \rangle \langle \tilde{I_0^2} \rangle}}
    \label{eq:correlation_formula}
\end{equation}
with $\tilde{I} = I - \langle  I \rangle $, and where $\langle \rangle$ stands for spatial averaging over all speckle grains.
The $0$ index represents the reference speckle image at $z=0$ and $\lambda_0$ = \SI{800}{nm}, to which all the other ones are compared.
Importantly, before computing the correlation product of experimental images we numerically corrected them from potential transverse drifts and intensity inhomogeneities~\cite{SM}.
The correlation products of so-corrected experimental intensity patterns are shown in~\fig{experimental_setup}d as a function of $\delta z$ and $\delta \lambda$.
%%FIGURE 1%%%%%%%%%%%%%%%%%%%%%%%%%%%%%%%%%%%%%%%%%%%%%%%%%%%%%%%
%%%%%%%%%%%%%%%%%%%%%%%%%%%%%%%%%%%%%%%%%%%%%%%%%%%%%%%%%%%%%%%%%

\begin{figure}[t]
\centering
    \includegraphics[width= 1\columnwidth]{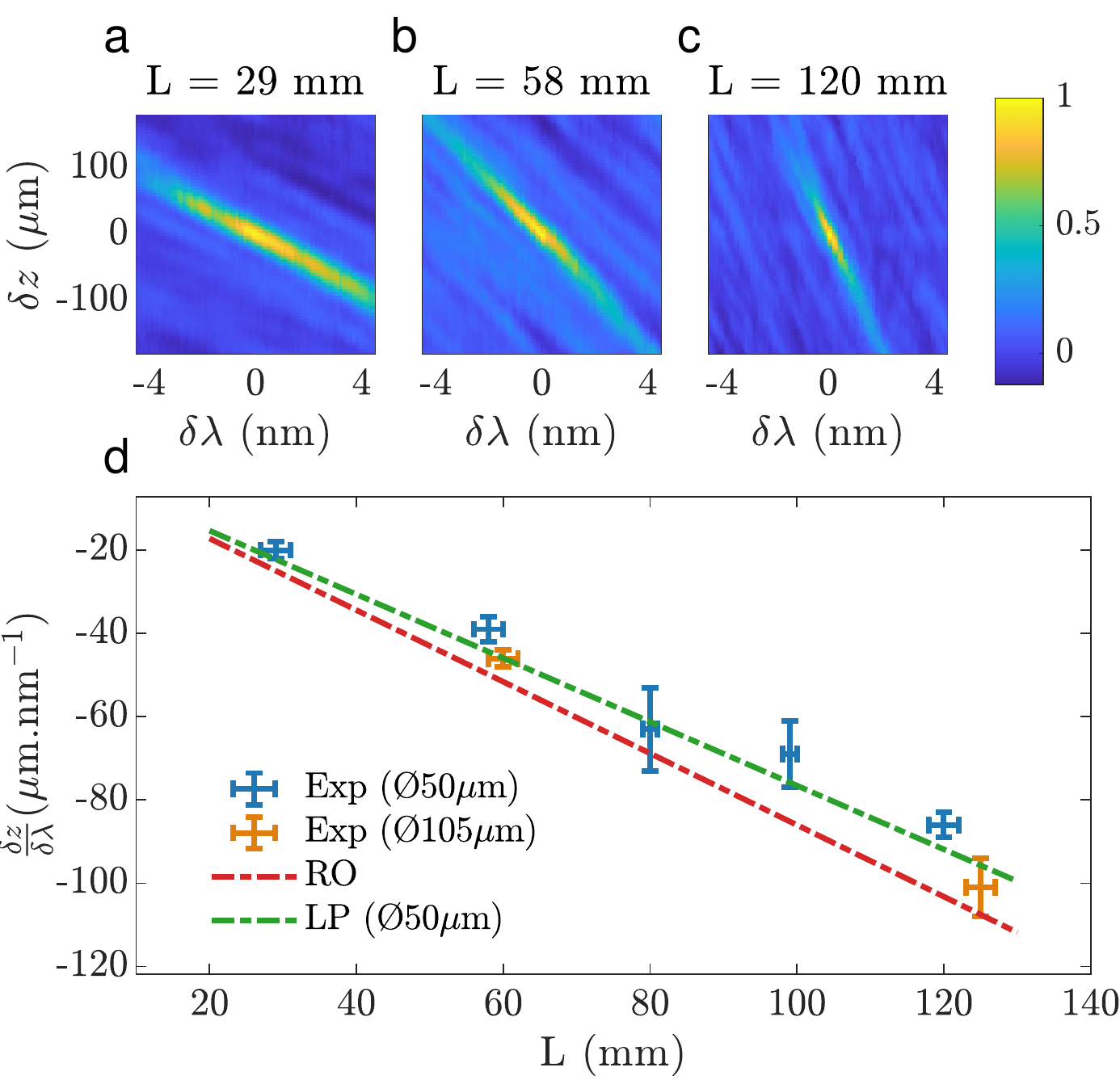}
    \caption{
    Correlations in the ($\delta z$, $\delta \lambda$) plane plotted for three fiber lengths (a) \SI{29}{\milli\metre} (b) \SI{56}{\milli\metre} (c) \SI{120}{\milli\metre}.
    The observed effect is present in all the three cases but its slope ($\delta z$/$\delta \lambda$) depends on the fiber length.
    (d) Experimental slopes measured for different fiber lengths and two core diameters (Ø = \SI{50}{\micro \metre}, blue dots, and Ø = \SI{105}{\micro \metre}, orange dots).
    The expected slope obtained from ray optics (RO) calculations is represented with the red dashed line.
    The green dashed line represents the slope obtained from the calculation with the LP modes (for Ø = \SI{50}{\micro \metre} and $\epsilon$ = 0.11, see~\Eq{Xaxial}).
    Error bars are obtained from 95\% confidence bounds of the slope extraction and fiber length measurement.
    }
    \label{fig:three_lengths}
\end{figure}

Similarly to what we obtain for a focused spot, a $\chi$-axial ME is obtained for output speckles: a correlation line appears when comparing speckle, proving the coupling between $\lambda$ and $z$ shifts.
In addition, these plots illustrate that the $\chi$-axial ME extends over limited spectral and axial ranges and that the longer the fiber the steeper the correlation slope (see~\Figs{three_lengths}a-c).
Noteworthy, the spectral and axial correlation ranges associated with the $\chi$-axial coupling are both larger than the ones usually defined for fixed $\lambda$ (Rayleigh length: $l_z = \frac{2 \lambda}{\rm NA^2}$~\cite{longitudinal_speckle}) and $z$ ($l_{\lambda} = \frac{2n_1 \lambda^2}{L \rm{NA}^2}$~\cite{rawson1980frequency,redding2013all}).
$l_z$ and $l_{\lambda}$ are illustrated in~\Fig{experimental_setup}d and their experimental values for our fibers are given and discussed in~\cite{SM}.
We thus now present a simple model allowing us to discuss how the correlation slope scales with physical parameters in the ray-optics approximation.

%%FIGURE 2%%%%%%%%%%%%%%%%%%%%%%%%%%%%%%%%%%%%%%%%%%%%%%%%%%%%%%%

%%%%%%%%%%%%%%%%%%%%%%%%%%%%%%%%%%%%%%%%%%%%%%%%%%%%%%%%%%%%%%%%%

 \medskip

Since the fiber diameters (\SI{50}{\micro \meter} and \SI{105}{\micro \meter}) are large in comparison to the working wavelength $\lambda_0$ = \SI{800}{\nano\metre}, we first consider a ray optics approximation model~\cite{Jackson}.
In a SI-fiber, the modulus of the transverse component of the wavevector is conserved, and the optical path length along a multiply bouncing trajectory scales quadratically with the incident angle~\cite{li2021memory}, similarly to the phase accumulated under free-space propagation in the Fresnel approximation.
Fresnel diffraction formula implies that the intensity is unchanged when the product $\lambda z$, appearing as the prefactor of the quadratic phase, is conserved, so resulting in the $\chi$-axial effect~\cite{zhu2020chromato}. 
In a SI-MMF, taking the
differential of the equation $\lambda z={\rm const.}$ results in the following slope for the $\chi$-axial ME at the fiber output:
\begin{equation}
    \frac{\delta z}{\delta \lambda} = -\frac{L}{n_1 \lambda_0},
    \label{eq:slope}
\end{equation}
where L is the fiber length, and where $n_1$, the refractive index contrast of the core with the output medium, is due to Snell-Descartes's law applying at the output facet of the MMF (see~\cite{SM} for a detailed derivation of~\Eq{slope}).
A spectral shift of the impinging beam thus results in a homothetic axial dilation of the intensity in the SI-MMF, with the origin of the homothety at the input facet of the fiber.
This is noticeably different from the $\chi$-axial ME in forward scattering slabs~\cite{vesga2019focusing} wherein the origin of the dilation was found to be in a virtual plane located at $1/3$ of the slab thickness~\cite{zhu2020chromato,Arjmand2021three}.
The slope obtained from~\Eq{slope} is plotted in~\fig{three_lengths}d, showing good agreement with the corresponding experimental measurements.
Noteworthy, within the framework of the ray optics model, the slope does not depend on the fiber radius, in qualitative agreement with experimental observations for the \SI{50}{\micro \meter} and the \SI{105}{\micro \meter} fibers that exhibit similar slopes, since this model is justified in both cases.
However, even though the ray model well predicts the measured slopes, it does not explain why the effect remains limited to finite $\chi$-axial ranges, as observed experimentally (\fig{experimental_setup}d and~\Figs{three_lengths}a-c).
A wave model is thus required to understand and estimate the limited spectral correlation width $\Delta \lambda$ and the limited axial range scanning ability $\Delta z$ that we observe in experiments.
In the following, we thus propose to analyse our results in the framework of the SI-MMF eigen-modes, assuming cylindrical symmetry for the fibers.

 \medskip

Here, we express the fibers modes in terms of linearly polarized (LP) modes, to then calculate an analytical expression for the cross correlation product (\Eq{correlation_formula}) as a function of $z$ and $\delta \lambda$.
The slope of the $\chi$-axial effect as well as its spectral and axial bandwidths arise from this expression.
Importantly, all calculations are carried out inside the fiber and do not consider free-space propagation outside, making these results not specific to our experimental system but revealing intrinsic spectro-axial properties of SI-MMF.
Also, for the sake of simplicity of expressions, spectral components of the field are now described by their angular frequencies $\omega=2\pi c/\lambda$, with $c$ the light speed in vacuum, rather than by their wavelength $\lambda$.

SI-MMF propagation eigenmodes are, in the weak guidance approximation, LP modes~\cite{snyder2012optical}.
The field propagating in a fiber of radius $a$ can then be expressed as:
\begin{equation}
E_{\omega}({\bf r},z)=\displaystyle{\sum_{l,m}} {\widetilde E}_{l,m}(\omega)e^{i\beta_{l,m}z}e^{il\varphi}J_l\left(u_{l,m}\frac{r}{a}\right)
\label{eq:field}
\end{equation}
where $l$ and $m$ are the azimuthal and the radial number of the LP-modes, respectively, $J_l$ the Bessel function of the first kind of order $l$, and where $\beta_{l,m}$ and $u_{l,m}/a$, the longitudinal and the transverse wavenumbers, are imposed by continuity equations at the core-cladding boundary.

Assuming that the fields in the fiber are random patterns with Gaussian statistics~\cite{reed1962moment}, the cross-correlation product used experimentally (\Eq{correlation_formula}), may be written as:
\begin{equation}
C(\omega,\omega^\prime,z,z^\prime) \propto |\left< E_{\omega} E_{\omega^\prime}^{\prime \ast}\right>|^2,
\label{eq:C1}
\end{equation}
up to a normalization factor. As discussed in \cite{SM}, intensity correlations at the fiber output are dominated by the phase delays accumulated along the fiber due to the chromatic dependence of $\beta_{l,m}$:
\begin{equation}
C(\omega,\omega^\prime,z,z^\prime)= \left| \left< e^{i\left[\beta_{l,m}(\omega)z-\beta_{l,m}^\prime (\omega^\prime) z^\prime\right]}\right>_{l,m} \right|^2.
\label{eq:normalized_CF}
\end{equation}
where, assuming ergodic hypothesis, statistical averaging is replaced by modal averaging over $l$ and $m$ values.
\Eq{normalized_CF} then appears as the characteristic function of variables $\delta\left[\beta_{l,m}(\omega)z\right]= \beta_{l,m}(\omega)z-\beta_{l,m}^\prime(\omega^\prime) z^\prime$, which we approximate to the Normal one, hence giving:
\begin{equation}
C(\omega,\omega^\prime,z,z^\prime) \simeq \exp\left(-\Var\left\{\delta[\beta_{l,m}(\omega)z]\right\}\right),
\label{eq:phase_variance}
\end{equation}
where the variance is thus calculated over all possible $l$ and $m$ values.
The axial shift $\delta z$ that maximizes the correlation coefficient (i.e. that minimizes the variance) for a given spectral shift $\delta \omega$ is thus an extremum of $C$ obtained by solving:
\begin{equation}
\frac{\partial}{\partial (\delta z)}\Var\left[\delta(\beta_{l,m}z)\right]=0
\label{eq:Xaxial_max}
\end{equation}
leading to (see \cite{SM}):
\begin{equation}
\delta z=(1-\epsilon)\frac{z}{\omega}\delta \omega,
\label{eq:Xaxial}
\end{equation}
where $\epsilon$ is a positive constant scaling as $1/v$, with $v = \omega a$NA/c, the normalized frequency.
For large fiber cores ($\omega a/c\gg 1$), $\epsilon$ thus vanishes. Numerical simulation further reveals that the product $v\epsilon$ does not strongly depend either of ${\rm NA}$ or of the core diameter.
On average $v\epsilon$ is found to be numerically on the order of 4.7, both for a \SI{50}{\micro \meter} or a \SI{105}{\micro \meter} core diameter~\cite{SM}.
In the large cores limit, \Eq{Xaxial} thus leads to \Eq{slope} obtained in the ray optic modeel.
The axial shift analytically found in~\Eq{Xaxial} in the frame of LP-mode modeling is compared to the ones obtained experimentally and theoretically in the ray-optics framework in~\Fig{three_lengths}d..
To compare~\Eq{Xaxial} to the experimental data, the slope value in~\Eq{Xaxial} is divided by $n_1$ to take into account Snell-Descartes relation when exiting the fiber.

%%FIGURE 3%%%%%%%%%%%%%%%%%%%%%%%%%%%%%%%%%%%%%%%%%%%%%%%%%%%%%%%
\begin{figure}[t]
\centering
    \includegraphics[width= 1\columnwidth]{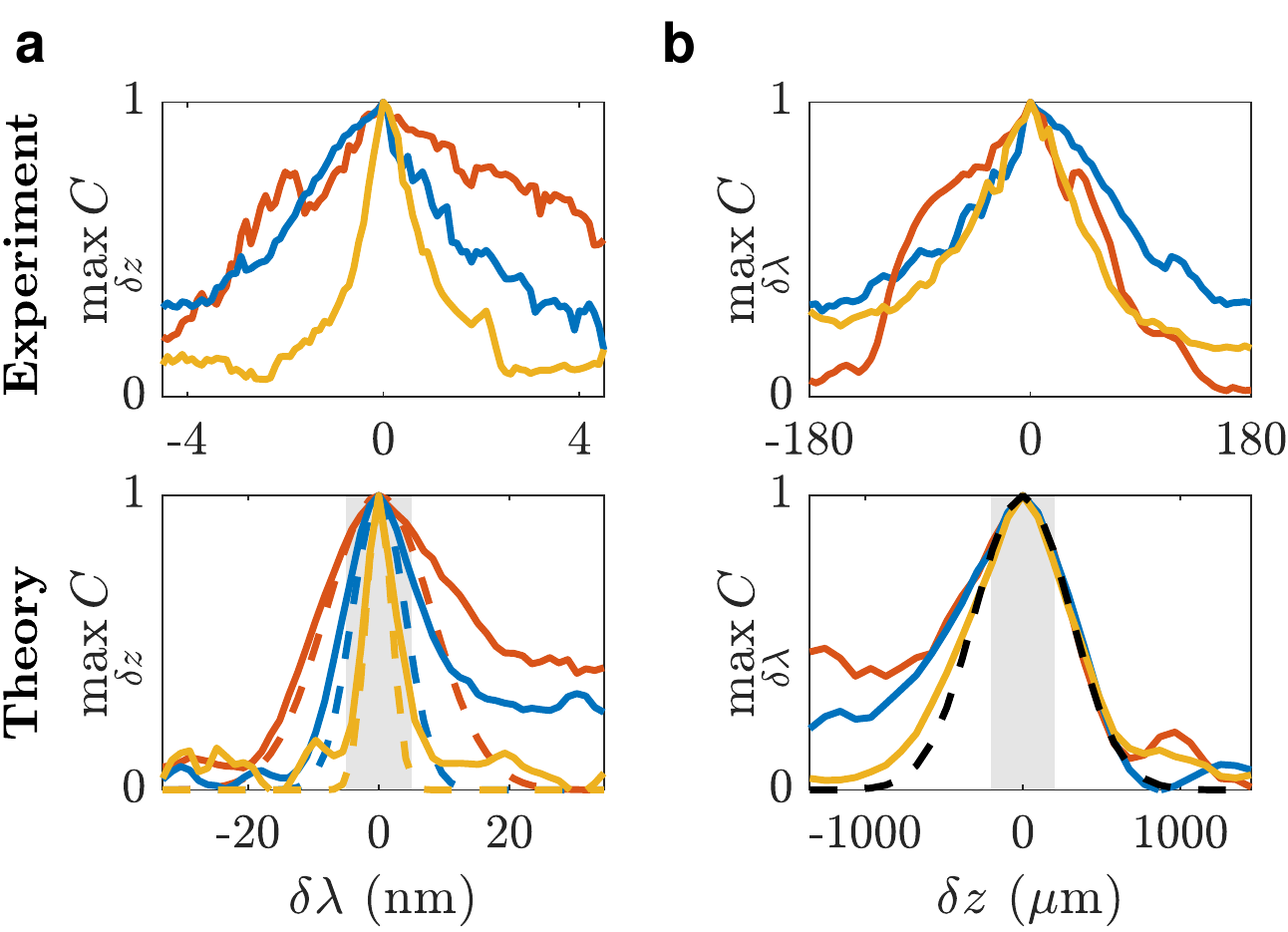}
    \caption{
    Ranges of the $\chi$-axial ME effect.
    Top graphs represent the experimental correlation plots projected on the $\lambda$ (resp. $z$) axis.
    Different fibres (of \SI{50}{\micro \meter} diameter) are distinguished by line colors: red for L = \SI{29}{\milli \meter}, blue for \SI{58}{\milli \meter} and yellow for \SI{120}{\milli \meter}.
    Bottom graph displays fiber intrinsic spectral and axial widths (solid lines) and their comparison to the theoretical values obtained from the LP model (dashed lines) for the same three fiber lengths.
    (a) Spectral width varies with the fiber length.
    The bottom gray domains indicate where the top corresponding experimental data are measured.
    (b) Axial width: because it does not depend on the fiber length only one plot (black dashed line) is presented for the analytical model.
    }
    \label{fig:spectral_width}
\end{figure}

%%%%%%%%%%%%%%%%%%%%%%%%%%%%%%%%%%%%%%%%%%%%%%%%%%%%%%%%%%%%%%%%%

 \medskip
 
We now consider the correlation width of the $\chi$-axial ME along the correlation line observed experimentally (\Figs{three_lengths}a-c).
Experimental spectral widths and axial displacement amplitudes are measured full width at half maximum (FWHM) by projecting maximal values of the cross-correlation function on the $\lambda$ and $z$ axes (see~\fig{spectral_width} top).
The spectral correlation widths of fibers are observed to depend on their lengths.
Conversely, the axial correlation width remains almost unchanged in relation to the fiber length.
We measure spectral widths of \SI{7}{\nano \meter}, \SI{3}{\nano \meter}, \SI{1}{\nano \meter} for the fibers of lengths \SI{29}{\milli \meter}, \SI{58}{\milli \meter} and \SI{120}{\milli \meter}, respectively, and a mean value of \SI{130}{\micro \meter} for the axial correlation width.

The analytical expression of the correlation width of the $\chi$-axial ME along the correlation line may be obtained by plugging~\Eq{Xaxial} into the expression of the correlation function \Eq{phase_variance}.
Replacing $\delta z$ by $\delta \omega$ in \Eq{phase_variance} according to \Eq{Xaxial}, we get:
\begin{equation}
\max_{\delta z} C=\exp\left[-\left(\frac{\delta \omega}{\Delta \omega}\right)^2\right]
\label{eq:equation9}
\end{equation}
with 
\begin{equation}
    \Delta \omega=\alpha \frac{2n_1\omega a}{{\rm NA}z},
    \label{eq:deltak}
\end{equation} 
where $\alpha$ is a prefactor close to one on average and whose exact value is discussed in~\cite{SM}.
The spectral width results are presented \Fig{spectral_width}a (bottom), where we plotted together the analytical formula \Eq{equation9} (dashed lines) and the results of numerical simulations (solid lines). 
As expected and experimentally observed the spectral width depends on the fiber length.
From the analytical LP model, we calculate expected spectral widths of \SI{20}{\nano \meter}, \SI{10}{\nano \meter} and \SI{5}{\nano \meter}, for fibers of lengths \SI{29}{\milli \meter}, \SI{58}{\milli \meter} and \SI{120}{\milli \meter}, respectively.
These theoretical values are thus larger than the experimental ones by a factor $\sim 3$ for the shortest fiber, up to $5$ for the longest fiber but both the orders of magnitude and the global evolution trend with fiber lengths agree.

Making use of~\Eq{Xaxial}, the correlation product can also be expressed as a function of the axial shift:
\begin{equation}
\max_{\delta \omega} C=\exp\left[-\left(\frac{\delta z}{\Delta z}\right)^2\right]
\label{eq:axial_width}
\end{equation}
with the axial correlation range
\begin{equation}
    \Delta z=\alpha(1-\epsilon) \frac{2n_1a}{{\rm NA}}.
\end{equation}
which, in agreement with experimental observations, does not depend on the fiber length.
The simulated axial width and the Gaussian analytical prediction are shown in~\Fig{spectral_width}b (bottom).
The axial correlation range obtained with the LP model is \SI{700}{\micro \meter} inside the fiber, yielding \SI{480}{\micro \meter} outside the fiber, once Snell-Descartes law is taken into account.
As for spectral correlation width, the theoretical value is a factor $3$ above the one obtained experimentally (\SI{130}{\micro \meter}).
In \Fig{spectral_width} it appears that the theoretical model matches the numerical simulations, except for the largest values of $\delta \lambda$ and the smallest values of $\delta z$.
We attribute this discrepancy to the limit of validity of our modeling hypothesis.

 \medskip

A detailed analysis of the contribution of the different modes to the total $\chi$-axial ME presented in~\cite{SM} reveals the key role of the less confined modes.
Our model assumes that the transverse profiles of LP modes is achromatic, but this approximation becomes inaccurate when the spectral correlation width is too large, as implied for short fibers; 
hence the better matching of the model with numerical simulations for the longest fiber than for the shortest one in~\Fig{spectral_width}.
The spectral sensitivity of the less confined modes also suggests that it is possible to broaden the spectral correlation width of the $\chi$-axial ME by limiting the light coupling to the lowest NA-modes, or by filtering high-NA modes at the output (i.e. working at a numerical aperture slightly below the one of the fiber, still allowing high-resolution imaging if a in-purpose higher-NA fiber is used).
Another way to improve the results could be to better fulfill the achromatic input field hypothesis by imaging the SLM in the fiber proximal end instead of imaging it on the microscope objective back aperture.

Finally, for applications where the $\chi$-axial ME is of interest after exiting the fiber, the intrinsic correlation width of the fiber is not the only limitation to consider: a mere geometric limitation arises.
Indeed when imaging planes far away, outside the fiber emission cone (defined by the fiber radius and its numerical aperture), the maximum spatial frequency decreases resulting in larger patterns that no-longer correlates with fields nearby the fiber output facet.
This geometrical effect thus limits the axial scan to $\delta z_c = \frac{2a}{\rm NA} \simeq \SI{230}{\micro \metre}$ that translates into a spectral width limitation adding to the intrinsic correlation range of the fiber.
Importantly for practical applications, both limitations are of same the order of magnitude: the FWHM of the intrinsic LP axial width is  (using~\Eq{axial_width}) $2 \ln{(2)} \Delta z \simeq 2.5 \delta z_c$. 
None of the mentioned phenomena should then dominate.

 \medskip

In conclusion, we experimentally demonstrated and quantified the $\chi$-axial ME in a SI-MMF.
Two different theoretical approaches enable us to predict the value of the $\chi$-axial shift.
A simple ray optics model gives access to the shift only using easily accessible fiber parameters with a good degree of approximation.
However, in the frame of this model, an infinite spectral width is predicted and slight discrepancies  arise on the slope of the $\chi$-axial ME when the field penetration inside the fiber cladding cannot be neglected.
The framework of LP modes not only brings more accuracy on the slope of the $\chi$-axial ME but also allows the derivation of its correlation width.
We observe that studied $\chi$-axial scanning ranges are not infinite and also obtain that it is possible to optimize the effect: analytical results demonstrate that working with the most confined modes would allow extending significantly the spectro-axial correlation range of this memory effect.
For imaging purposes, when a large spectral bandwidth is required, working with short MMF should be preferred.
Alternatively, avoiding the excitation of the less confined modes is expected to increase the spectral correlation range, in principle up to arbitrarily large values.

Characterization of the $\chi$-axial ME in MMFs is of high importance to simplify light-beam manipulation by blind control of polychromatic wavefields at distal facets, especially for imaging applications.
Thanks to the $\chi$-axial ME, only one TM measurement would be needed to perform foci on different z planes, and thereby overcome the current inability of wavelength tuning while imaging~\cite{pikalek2019wavelength}.
Furthermore, the $\chi$-axial ME knowledge enables \emph{a priori} wavefront correction to achieve non-linear microscopy~\cite{morales2015two,Brasselet_PRAppl_20}, and to extend the confocal microscopy technique of~\cite{loterie2015confocal} for objects subjected to inelastic scattering or broadband fluorescence.
This fast axial scan ability opens up the possibility of extending the “spot scanning” imaging technique in three dimensions~\cite{papadopoulos2012focusing},
paving the way to non-invasive imaging (in biological media for instance). 

\section{Acknowledgments}
We would like to thank Micha\l{} D\k{a}browski and Lorenzo Valzania for the careful reading of the manuscript and fruitful comments.
This project was funding by the European Research Council under the grant agreement No. 724473 (SMARTIES).
SG and MG are members of the Institut Universitaire de France.

\bibliography{Paper_SI_chromato_axial}

%%%%%%%%%%%%%%%%%%%%%%%%%%%%%%%%%%%%%%%%%%%%
%%%%%%%%%% Supplemental Materials %%%%%%%%%%
%%%%%%%%%%%%%%%%%%%%%%%%%%%%%%%%%%%%%%%%%%%%

\clearpage
\onecolumngrid

\renewcommand{\thefigure}{S\arabic{figure}}
\renewcommand{\theequation}{S\arabic{equation}}
\setcounter{equation}{0}
\setcounter{figure}{0}

\begin{center}
  \LARGE
  \textbf{Supplemental Materials}
\end{center}

\section{Material}

In the experimental setup presented~\fig{experimental_setup}a of the main text, the monochromatic tunable laser is a MaiTai HP (Spectra Physics).
The CCD camera is a Basler ace (acA1300-30uc) and the SLM is from Meadowlarks (HSP512L-1064).
Also the MMFs used for the experiments are purchased from Thorlabs (FG050LGA/FG105LCA).

\section{Spectral widths at fixed $\lambda$ and $z$}

The axial correlation extent $l_z$ of a monochromatic speckle equals the Rayleigh length: $l_z = \frac{2 \lambda}{\rm NA^2}$~\cite{longitudinal_speckle}.
From our experimental measurements presented in~\Fig{experimental_setup}d and \Figs{three_lengths}a-c of the main text, we obtain $l_z \simeq \SI{31}{\micro \meter}$ FWHM, in agreement with the Rayleigh length expression ($l_z$ = \SI{33}{\micro \metre} for a $0.22$-NA fiber at \SI{800}{\nano \metre}).
Correspondingly, the spectral width $l_{\lambda}$ in a given transverse plane, is $l_{\lambda} = \frac{2n_1 \lambda^2}{L \rm{NA}^2}$~\cite{rawson1980frequency,redding2013all}.
We measured spectral widths
(FWHM) equal to 1.4, 0.8 and \SI{0.4}{\nano \meter} for L = 29, 58 and \SI{120}{\milli \metre}, respectively, in agreement with the analytical expression of $l_{\lambda}$.
Both $l_z$ and $l_{\lambda}$ lengths are represented in \Fig{experimental_setup}d.

\section{Experimental precautions and data processing}

\textbf{Alignment post-correction}:
To prevent any correlation drop due to unavoidable alignment issues during the z-scan we used a reference.
For each experiment we performed a z-scan with a white light beam and tracked its center.
The center displacement enabled us, by cropping the speckle data, to correctly correlate the images and minimize any misalignment induced correlation drop.

\medskip

\begin{figure}[h!]
    \includegraphics[width=0.8\columnwidth]{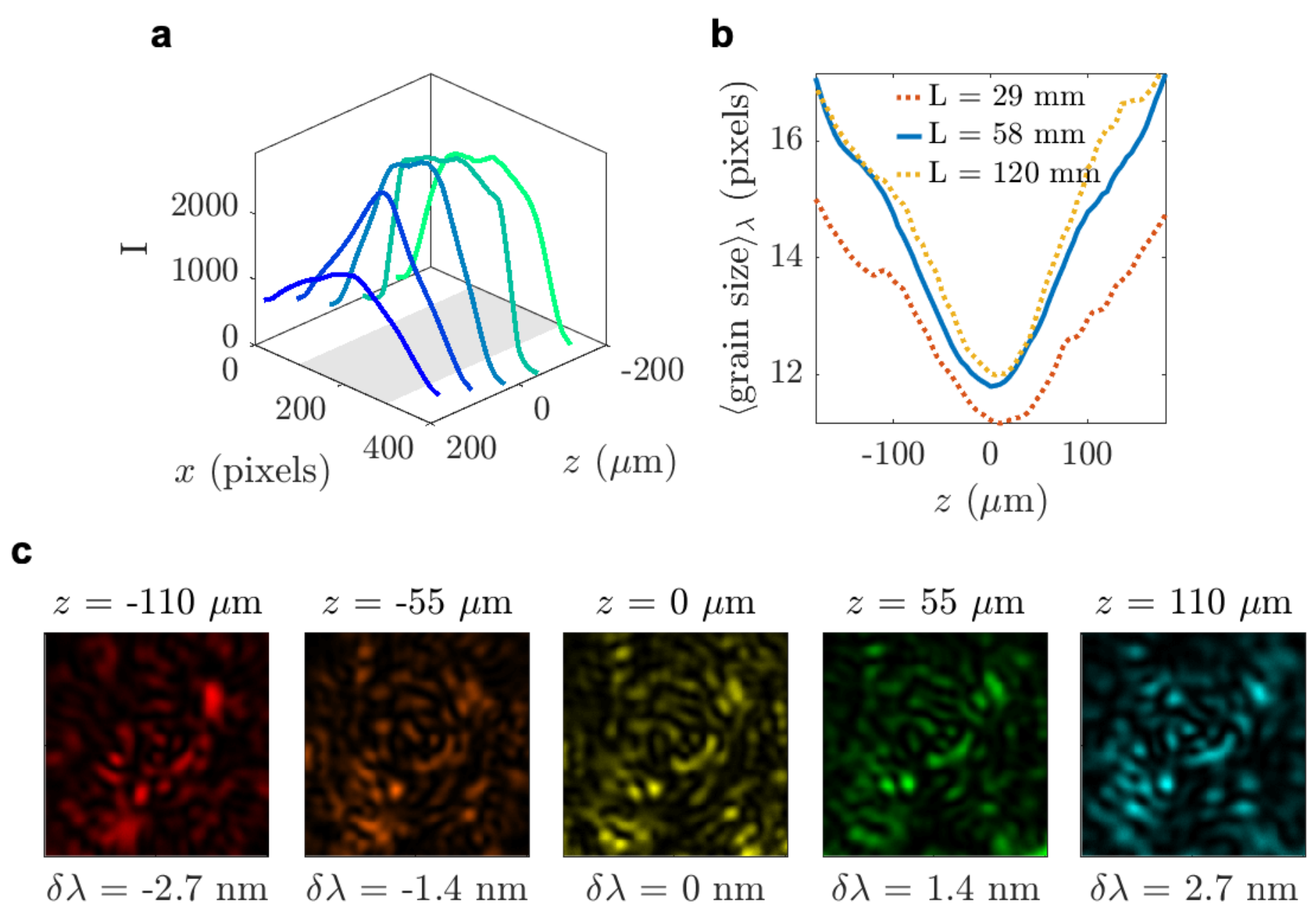}
    \caption{
    (a) Evolution of the white light intensity profile along the z axis for a cut on the y dimension for the \SI{58}{mm} long SI fiber.
    The bottom gray mark represents the speckle central position kept to compute the correlations.
    (b) Evolution of the speckle grain size along the z axis.
    The solid blue line represents the mean grain size (mean over all wavelength shifts) for the L = \SI{58}{\milli \meter} fiber.
    The relative variation is $R_{58}$ = 0.45.
    The red (resp. yellow) dotted lines is the mean grain size evolution for L = \SI{29}{\milli \meter} (resp. \SI{120}{\milli \meter}) fiber giving a relative variation of $R_{29}$ = 0.34 (resp. $R_{120}$ = 0.45).
    The mean relative variation for these three experiments is $\langle R \rangle$ = 0.42.
    (c) Speckles of raw images along the $\chi$-axial high correlation line.
    No important transverse dilatation is observed.
    }
    \label{fig:SM_zscan}
\end{figure}

\textbf{Correlation measurement}:
Before computing the correlations we selected on the data the central region with approximately 200 speckle grains.
In this region we used the reference white light z-scan to measure an intensity profile for each axial position (see~\Fig{SM_zscan}a).
Indeed if the profile is flat for $z=0$, imaging planes away from this position leads to intensities peaked at the centre.
We then used the measured reference intensity profiles to correct the speckle images.
The reason of these two precautions is the following: if, in addition to the speckle short range variations, long range variations (due to the spatial illumination non-flatness) are present, they impact the correlation value.
One hence may observe a -- relatively important -- correlation background and the correlation value does not reach zero for uncorrelated speckles.
It is however important to point out that the $\chi$-axial effect is still well visible in absence of correction and that the later only affect the correlation background.

\medskip

\textbf{Speckle dilation and z-scan}:
For planes away from the fiber output ($z = 0$), the imaged speckle is dilated due to Fresnel diffraction.
To evaluate the dilation we measured the speckle grain size by performing the autocorrelation of the intensity image and extracting its FWHM.
The~\Fig{SM_zscan}b shows the speckle grain size evolution along z for the L = \SI{58}{\milli \meter} fiber. For each z position the grain size value results from the average for all wavelengths.
For this experiment the relative variation of grain size is of $R_{58}$ = 0.45.
The mean value for all three main experiments is $\langle R \rangle$ = 0.42.\\
However because of free space propagation the speckle transverse dilatation on the axial scan is compensated by the wavelength detuning such that on the high correlation line of the effect no dilatation is observed as visible on~\Fig{SM_zscan}c.

%%% Fig. SM1?  %%%%%%%%%%%%%%%%%%%%%%%%%%%%%%%%%%%%%%%%
%%%%%%%%%%%%%%%%%%%%%%%%%%%%%%%%%%%%%%%%%%%%%%%%%%%

\medskip

\textbf{SLM impact}:
We also checked that the chromaticity of the SLM had no impact on the experimental results by suppressing it on one experiment.
It did not affect the $\chi$-axial ME effect itself neither its quantitative value.

\section{Comparison SI-MMF, GI-MMF and lens chromaticity}

%%% Fig. SM1  %%%%%%%%%%%%%%%%%%%%%%%%%%%%%%%%%%%%%%%%
\begin{figure}[h!]
    \includegraphics[width=0.4\columnwidth]{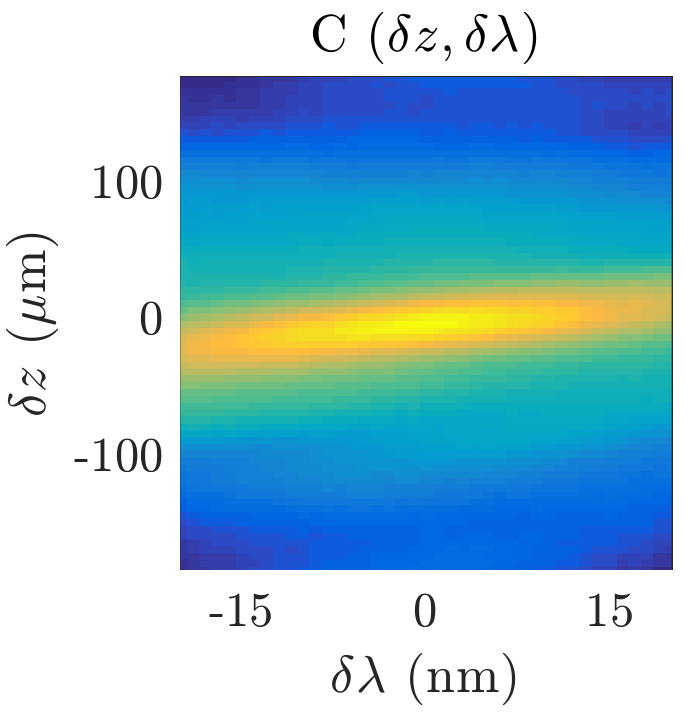}
    \caption{Correlation plot in the ($\delta z$, $\delta \lambda$) axis for a \SI{58}{mm} length graded index fiber of 0.22-NA.
    One observe a slope of $\simeq$ \SI{0.5}{\micro \meter \per \nano \meter}
    The order of magnitude as well as the sign of the shift are different with what would be expected for the $\chi$-axial effect presented in the main text.
    }
    \label{fig:SM_GI_fibre}
\end{figure}
%%%%%%%%%%%%%%%%%%%%%%%%%%%%%%%%%%%%%%%%%%%%%%%%%%%

To ensure that the $\chi$-axial shift is not a trivial effect we evaluated the shift one could expect from the mere chromaticity of the experimental setup optical components.
The microscope objectives used are treated against chromaticity in the visible range. At \SI{800}{\nano \meter} it remains $\frac{\delta z}{\delta \lambda} \approx$ \SI{0.4}{\micro \meter \per \nano \meter}.
The lens used is plane-convex, made of N.BK7 glass.
The chromaticity evaluated at \SI{800}{\nano \meter} gives $\frac{\delta z}{\delta \lambda} \approx$ \SI{0.1}{\micro \meter \per \nano \meter}.
These values are far different from the effect observed both in order of magnitude and in sign.
However it matches the slope obtained when performing the experiment with a GI fiber (see~\fig{SM_GI_fibre}).

\section{Calculations for ray optics model}

To derive the $\chi$-shift in the framework of ray optics, let us focus on the electric field on different planes perpendicular to the optical axis when the incident wavelength varies.
We make the following assumptions:
$E_{\omega}(x,y,-L)$ does not depend on the angular frequency $\omega$ (save for a global phase factor);
on the whole range of explored wavelengths, the refraction indices of the fiber ($n_1$ and $n_2$) do not depend on $\omega$;
the angles between optical rays and z-axis are small such that the paraxial approximation holds;
with a step-index fiber, $k_{\perp}^2$ and $k_z$ are conserved throughout the propagation inside the fiber, hence their input and output values are equal.
Hence the light rays follow the same path inside the fiber whatever the angular frequency: varying $\omega$ just implements an homothetic transformation of the wave-vector distribution, enabling to write:
\begin{equation}
\tilde{E}_{\omega}(k_x,k_y,z = 0) = \tilde{E}_{\omega_0}\left(\frac{\omega_0}{\omega}k_x,\frac{\omega_0}{\omega}k_y,z = 0\right) e^{i \Delta \phi},
\label{eq:SM_hypo}
\end{equation}
where $\Delta \phi = \Delta \phi (k_x, k_y, \omega, \omega_0) = \phi_{\omega} - \phi_{\omega_0}$ with $\phi$ the phase accumulated while propagating inside the fiber, and $\tilde{E}$ indicates the spatial Fourier transform of $E$.
Now expressing the electric field at a position $z$  for an angular frequency $\omega$ gives
\begin{align}
E_{\omega}(x,y,z) & = \int \int dk_x dk_y e^{i(k_x x + k_y y)} \tilde{E}_{\omega}(k_x, k_y, z) \nonumber \\
			& = \int \int dk_x dk_y e^{i(k_x x + k_y y)} \tilde{E}_{\omega}(k_x, k_y, 0)e^{ik_z z}.
\end{align}
where $k_z$ is derived from ($\omega$, $k_x$, $k_y$) through the dispersion relation : $\frac{\omega^2}{c^2} = k_x^2 + k_y^2 + k_z^2$ leading to $k_z \approx \frac{\omega}{c}\left(1-\frac{1}{2}\frac{c^2}{\omega^2}(k_x^2 + k_y^2)\right)$. Hence
\begin{equation}
E_{\omega}(x,y,z) = \int \int dk_x dk_y e^{i(k_x x + k_y y)} \tilde{E}_{\omega}(k_x, k_y, 0)e^{iz\frac{\omega}{c}\left(1-\frac{1}{2}\frac{c^2}{\omega^2}(k_x^2 + k_y^2)\right)}.\\
\end{equation}
Using~\Eq{SM_hypo} we obtain
\begin{equation}
E_{\omega}(x,y,z) = \int \int dk_x dk_y e^{i(k_x x + k_y y)} \tilde{E}_{\omega_0}\left(\frac{\omega_0}{\omega}k_x, \frac{\omega_0}{\omega}k_y, 0\right)e^{i \Delta \phi} e^{iz\frac{\omega}{c}}e^{-iz\frac{c}{2\omega}(k_x^2 + k_y^2)}.\\
\end{equation}
One needs to express $\Delta \phi$. We assume (with \textit{a posteriori} confirmation) that $\Delta \phi(k_x, k_y, \omega, \omega_0) = \Delta \phi_0(\omega, \omega_0) + \frac{k_x^2 + k_y^2}{(\omega/c)^2} \Delta \zeta(\omega, \omega_0)$.
This is equivalent to show that the phase $\phi$ is proportional/has a term in $\sin^2\theta$, where $\theta$ is the incident angle of the input light ray.
At this step the $k_{\perp}$ conservation is mandatory to identify the $k_{\perp}^2 = k_x^2 + k_y^2$ term in $\phi$ (therefore after propagation in the fiber) with the input $k_{\perp}^2 = \frac{\omega^2}{c^2}\sin^2 \theta$.
\begin{align}
E_{\omega}(x,y,z) & = e^{iz\frac{\omega}{c}} e^{i \Delta \phi_0} \int \int dk_x dk_y e^{i(k_x x + k_y y)} \tilde{E}_{\omega_0}\left(\frac{\omega_0}{\omega}k_x, \frac{\omega_0}{\omega}k_y, 0\right)e^{-i\frac{c}{2\omega}(k_x^2 + k_y^2)(z-2 \frac{\Delta \zeta}{\omega/c})} \nonumber \\
			& =  e^{iz\frac{\omega}{c}} e^{i \Delta \phi_0} \left (\frac{\omega}{\omega_0}\right) ^2 \int \int dk'_x dk'_y e^{i\frac{\omega}{\omega_0}(k'_x x + k'_y y)} \tilde{E}_{\omega_0}(k'_x, k'_y, 0)e^{-i\frac{c}{2 \omega} (\frac{\omega}{\omega_0})^2 (k'^2_x + k'^2_y)(z-2 \frac{\Delta \zeta}{\omega/c})}.
\end{align}
As a consequence, one has
\begin{equation}
E_{\omega}\left(x,y,z = \frac{2 \Delta \zeta}{\omega/c}\right) = A e^{i \psi} E_{\omega_0}\left(\frac{\omega}{\omega_0}x,\frac{\omega}{\omega_0}y, z =0\right).
\end{equation}
Thus, save for a spatial transverse magnification by a factor of $\frac{\omega_0}{\omega}$, both speckle $E_{\omega}$ resp. $E_{\omega_0}$ are proportional when considered at abscissa $z = \frac{2 \Delta \zeta}{\omega/c}$ resp. $z=0$.
It is noteworthy that for $\lambda_0$ = \SI{800}{\nano \meter} and $\Delta \lambda = 10-20 \, \rm{nm}$, the relative transverse magnification of $10^{-2}$, $2.10^{-2}$ is hardly observable.
If one limits to a short range of $\omega$, $\frac{\Delta \zeta}{\omega/c} = \frac{\zeta(\omega) - \zeta(\omega_0)}{\omega - \omega_0} \frac{\omega - \omega_0}{\omega/c} \approx \left. \frac{d \zeta}{d \omega}\right|_{\omega_0} d \omega \frac{1}{\omega_0/c}$.
So that,
\begin{equation}
\frac{dz}{d \omega} \approx \frac{2c}{\omega_0} \left. \frac{d \zeta}{d \omega} \right|_{\omega_0}.
\label{eq:SM_slope}
\end{equation}
 Note that a local slope $\frac{dz}{d \omega}$ also exists if $\omega$ is far from $\omega_0$.
 It is just slightly more cumbersome to write down its analytical value.
 There is no $\omega$-range limitation for this effect on this model.

To go forward, one needs to evaluate the phase $\phi_0$ accumulated during the light propagation in the fiber. 
For a pure ray model the accumulated phase during propagation is 
\begin{equation}
\phi_0 = \frac{L n_1 (\omega/c)}{\sqrt{1 - (\frac{\sin \theta}{n_1})^2}}.
\end{equation}
At small angles: $\phi_0 \approx  Ln_1\frac{\omega}{c}(1 + \frac{\sin^2 \theta}{2n_1^2})$, such that
\begin{equation}
\zeta = \frac{L \omega}{2 n_1 c}.
\label{eq:zeta_simple}
\end{equation}
Using~\Eq{SM_slope}, one gets
\begin{equation}
\frac{dz}{d \lambda} = \frac{-L}{n_1 \lambda_0}.
\end{equation}
As commented in the main text the value of the slope is in global agreement with the experimental data.

\section{Some complements for LP modes calculations}
\subsection{Derivation of~\Eq{normalized_CF} of the main text}

In this section we present the full derivation of the LP model used in the main text.
We first recall the derivation central thread.
After expressing field propagation in the fiber in terms of linearly polarized (LP) modes, we calculate an analytical expression for the cross correlation product (\Eq{correlation_formula}) as a function of $z$ and $\delta \lambda$ and deduce from this expression the slope of the $\chi$-axial effect as well as its spectral and axial bandwidths.
Along the derivation we point out correspondences between the assumptions used and the ray optics model.
Importantly, all the following calculations are carried out inside the fiber and do not consider free-space propagation outside, making these results not specific to our experimental system but revealing intrinsic spectro-axial properties of SI-MMF.
For the sake of simplicity of expressions, spectral components of the field are described by their angular frequencies (independent of the refractive index of the fiber-core) $\omega=2\pi c/\lambda$, with $c$ the light velocity in vacuum, rather than by their wavelength $\lambda$.

SI-MMF propagation eigenmodes are, in the weak guidance approximation, LP modes~\cite{snyder2012optical}.
The field propagating in a fiber of radius $a$ can then be expressed as:
\begin{equation}
E_{\omega}({\bf r},z)=\displaystyle{\sum_{l,m}} {\widetilde E}_{l,m}(\omega)e^{i\beta_{l,m}z}e^{il\varphi}J_l\left(u_{l,m}\frac{r}{a}\right)
\label{eq:SM_field}
\end{equation}
where $l$ and $m$ are the azimuthal and the radial number of the LP-modes, respectively, $J_l$ the Bessel function of the first kind of order $l$, and where $\beta_{l,m}$ and $u_{l,m}$, the longitudinal and the transverse wavenumbers, are imposed by continuity equations at the core-cladding boundary.
In particular, they satisfy the following equation:
\begin{equation}
 (\beta_{l,m}a)^2 + u_{l,m}^2 = \left(\frac{n_1 \omega a}{c}\right)^2.
\end{equation}
For low-NA fibers, a second order Taylor expansion of the former equation yields a simpler expression for $\beta_{l,m}$:
\begin{equation}
\beta_{l,m}\simeq \frac{n_1 \omega}{c} \left[1-\frac{1}{2}\left(\frac{u_{l,m} c}{n_1 \omega a} \right)^2\right].
\label{eq:SM_betasimplified}
\end{equation}

Assuming that the fields in the fiber are random patterns with Gaussian statistics~\cite{reed1962moment}, the cross-correlation product used experimentally (\Eq{correlation_formula} of the main text), may be written as:
\begin{equation}
C(\omega,\omega^\prime,z,z^\prime) \propto |\left< E_{\omega} E_{\omega^\prime}^{\prime \ast}\right>|^2,
\label{eq:SM_C1}
\end{equation}
up to a normalization factor.
Using \Eq{SM_field}, we get the expression of the two-wavelength mutual coherence function:
\begin{equation}
\left< E_{\omega} E_{\omega^\prime}^{\prime \ast}\right> =   \displaystyle{\sum_{l,l^\prime,m,m^\prime}} \left< {\widetilde E}_{l,m} {\widetilde E}_{l^\prime,m^\prime}^{\prime \ast}  e^{i(\beta_{l,m}z-\beta_{l^\prime,m^\prime}^\prime z^\prime)}\right. 
 \left. \times e^{i(l-l^\prime)\varphi} J_l\left(u_{l,m}\frac{r}{a}\right) J_{l^\prime}\left(u_{l^\prime,m^\prime}^\prime\frac{r}{a} \right)\right>,
\label{eq:SM_C}
\end{equation}

To calculate analytically Eqs.~\eqref{eq:SM_C1} and~\eqref{eq:SM_C}, we need to make a few simplifications.
Indeed, since $u_{l,m}$ depend on the wavelength, so do the LP-modes. 
When sending a field for which both amplitude and phase are assumed to be achromatic inside the fiber (for instance a pinhole illuminated with white light), the propagation of the different wavelengths should be written using different sets of eigenmodes. 
Alternatively, we shall consider that the transfer matrix from the eigen-basis at $\omega^\prime$ to the one at $\omega$ is not identity and thus, that the projection coefficients, for an achromatic impinging field, depend on the wavelength.
However, in practice, the transverse profiles $J_l\left(u_{l,m}\frac{r}{a}\right)$ only weakly depend on the wavelength because $u_{l,m}$ is close to the $m^{th}$ zero of $J_l$ and spectral detuning dependence of $u_{l,m}$ is a second order perturbation.
The following assumptions are then made:
\begin{itemize}
\item Spectral fluctuations of the $\widetilde {E}_{l,m}$ coefficients do not significantly contribute to the cross-correlation product (\Eq{SM_C1}), since they just modify the weighting.
Equivalently, although Bessel functions are not exactly orthogonal in~\Eq{SM_C}, their inner product will only contribute for a weighting change. 
\item In contrast, intensity correlations at the fiber output are dominated by the phase delays accumulated along the fiber due to the chromatic dependence of $\beta_{l,m}$.
\end{itemize}
The two-wavelength mutual coherence function then becomes,
\begin{equation}
\left< E_k E_{k^\prime}^{\prime \ast}\right> =  \pi a^2 \displaystyle{\sum_{l,m}} \left< {\widetilde E}_{l,m} {\widetilde E}_{l,m}^{\prime \ast} \left[J_{l+1}(u_{l,m}) \right]^2 e^{i(\beta_{l,m}z-\beta_{l^\prime,m^\prime}^\prime z^\prime)}\right>.
\end{equation}
It comes that $C$ is 
as a weighted average of phasors $e^{i(\beta_{l,m}z-\beta_{l^\prime,m^\prime}z^\prime)}$.
Assuming that the coefficients $\left\{\widetilde{E}_{l,m}\right\}_{(l,m)}$ are random independent variables (independent from $\left\{\beta_{l,m}\right\}_{(l,m)}$ in particular), the normalized correlation function is given by~\Eq{normalized_CF} of the main text, recalled below,
\begin{equation}
C(\omega,\omega^\prime,z,z^\prime)= \left| \left< e^{i\left[\beta_{l,m}(\omega)z-\beta_{l,m}^\prime (\omega^\prime) z^\prime\right]}\right>_{l,m} \right|^2.
\label{eq:SM_normalized_CF}
\end{equation}
As stated in the main text, we assume ergodic hypothesis making statistical averaging equivalent to modal averaging over $l$ and $m$ values.
\Eq{SM_normalized_CF} is the characteristic function of the variable $\delta\left[\beta_{l,m}(\omega)z\right]= \beta_{l,m}(\omega)z-\beta_{l,m}^\prime(\omega^\prime) z^\prime$. 
In order to obtain a closed form expression, we approximate the characteristic function of the modal distribution of $\delta\left[\beta_{l,m}(\omega)z\right]$ by a Normal one:
\begin{equation}
C(\omega,\omega^\prime,z,z^\prime) \simeq \exp\left(-\Var\left\{\delta[\beta_{l,m}(\omega)z]\right\}\right),
\label{eq:phase_variance_SM}
\end{equation}

Differentiation of $\delta\left[\beta_{l,m}(\omega)z\right]$, relying on~\Eq{SM_betasimplified} yields:
\begin{equation}
\delta \left(\beta_{l,m}z\right)= \delta\left(\frac{n_1\omega z}{c}\right)-\frac{c }{2n_1 (\omega a)^2} u_{l,m}^2\left[ \omega \delta z-(1-\chi_{l,m})z\delta \omega \right]
\label{eq:differentiation}
\end{equation}
with 
\begin{equation}
    \chi_{l,m}=2\omega\frac{\partial}{\partial \omega}\left[\ln(u_{l,m})\right],
\end{equation}
\Eq{differentiation} can then be written as a second order polynomial in $\omega\delta z$ and $z\delta\omega$:
\begin{equation}
    \Var\left\{\delta[\beta_{l,m}(\omega)z]\right\} = \left[\frac{c}{2n_1(\omega a)^2}\right]^2 \left[ A (\omega\delta z )^2 + 2 B(\omega\delta z )(z\delta\omega) + C(z\delta\omega)^2 \right]
    \label{eq:variance_ABC_SM}
\end{equation}
where $A$, $B$ and $C$ are given by:
\begin{align}
A & = \Var(u^2)\nonumber\\
B & = -\Var(u^2)+\left<\chi u^4\right>-\left<u^2\right>\left<\chi u^2\right> \label{eq:ABC_values} \\
C & = \Var[u^2(1-\chi)]\nonumber
\end{align}
where indices $l$ and $m$ have been dropped for the sake of brevity. The variance in Eq.~\eqref{eq:variance_ABC_SM} is thus the sum of two parabolas. The orthogonal eigen-axes of these parabola can be identified by writing Eq.~\eqref{eq:variance_ABC_SM} in a matrix form:
\begin{equation}
    \left[\frac{2n_1(\omega a)^2}{c}\right]^2\Var\left\{\delta[\beta_{l,m}(\omega)z]\right\} = \begin{pmatrix} \omega\delta z && z\delta\omega \end{pmatrix} \begin{pmatrix} A && B \\ B && C\end{pmatrix} \begin{pmatrix} \omega\delta z \\ z\delta\omega \end{pmatrix}
    \label{eq:variance_ABC_matrix_SM}
\end{equation}
where it appears that eigen-axes are eigen-vectors of the involved  $2\times 2$ matrix. Eigen-values are trivially:
\begin{equation}
    s_\pm = \frac{A+C\pm \sqrt{(A+C)^2-4(AC-B^2)}}{2}
    \label{eq:eigen_values_SM}
\end{equation}
Importantly, from Eqs.~\eqref{eq:ABC_values}, it appears that the determinant $AC-B^2$ appearing in Eq.~\eqref{eq:eigen_values_SM} vanishes if $\chi$ has a Dirac distribution. In practice, the distribution of $\chi$ is highly peaked and deviation from the peak value is due to the less confined modes as illustrated in Fig.\ref{fig:SM_chi}.
Since $u$ scale as $v$ and $\chi$ as $1/v$ (see~\Eq{dudk}), with $v = \omega a\rm{NA}/c$, the normalized frequency, it is easy to show that $AC-B^2$ scales as $v^6$ while $(A+C)^2$ scales as $v^8$.
Consequently, for highly multimode fibers, for which $v\gg 1$, we may simplify the expressions of eigen-values $s_\pm$:
\begin{align}
    s_+ & \simeq A+C\\
    s_- & \simeq \frac{AC-B^2}{A+C} 
\end{align}
resulting in a ratio $\frac{s_+}{s_-}\simeq\frac{(A+C)^2}{AC-B^2}$ scaling as $v^2\gg 1$. Relying on the same approximation, the expression of the eigen-vectors are then 
\begin{align}
    V_+ & \simeq \frac{1}{\sqrt{A^2+B^2}} \begin{pmatrix} A \\ B \end{pmatrix}\\
    V_- & \simeq \frac{1}{\sqrt{A^2+B^2}} \begin{pmatrix} -B \\ A \end{pmatrix}
\end{align}
In the eigen-basis $(V_+,V_-)$ 
of the system, the correlation function in Eq.~\eqref{eq:phase_variance_SM} can thus be simplified and expressed in the eigen-system of coordinates $(\xi_+,\xi_-)$ as:
\begin{equation}
C(\xi_+,\xi_-) \simeq \exp\left\{- \left[\frac{c}{2n_1(\omega a)^2}\right]^2(s_+ \xi_+^2 + s_- \xi_-^2) \right\}.
\label{eq:correlation_eigenbasis_SM}
\end{equation}
with:
\begin{align}
    \xi_+ &=V_+\cdot \begin{pmatrix} \omega\delta z \\ z\delta\omega \end{pmatrix} \simeq \frac{A\omega\delta z + Bz\delta \omega}{\sqrt{A^2+B^2}}\\
    \xi_- &=V_-\cdot \begin{pmatrix} \omega\delta z \\ z\delta\omega \end{pmatrix} \simeq \frac{-B\omega\delta z + Az\delta \omega}{\sqrt{A^2+B^2}}
\end{align}
As a result, the sought-for correlation (Eq.~\eqref{eq:phase_variance_SM}) is a product of two Gaussians, one of which being much larger than the other because $s_+\gg s_-$, so resulting in a cigar-like shape of the correlation function.
It is noteworthy that the $\chi$-axial ME is well established by diagonalizing the correlation function.
This approach is similar to~\cite{li2021memory} and shows that not only diagonalizing the TM brings information on the field transformation but that the speckle correlation function, sometimes quite easily analytically accessible, also does.
Indeed the intensity cross-correlation was already shown to allow interferometric measurements~\cite{freund1990looking}.

\subsection{Expression of $\epsilon$}
The long axis of the cigar-shaped correlation function is a line satisfying $\xi_+=0$, so yielding~\Eq{Xaxial_max} in the main text:
\begin{align}
\delta z&=-\frac{B}{A} \frac{z}{\omega}\delta \omega \label{eq:SM_Xaxial_a}\\
& = (1-\epsilon)\frac{z}{\omega}\delta \omega,
\label{eq:SM_Xaxial}
\end{align}
where $\epsilon$ is:
\begin{equation}
\epsilon = \frac{ \left<\chi u^4\right>-\left<u^2\right>\left<\chi u^2\right>}{\Var(u^2)},
\label{eq:SM_epsilon}
\end{equation}
From numerical simulation, we observe that $\epsilon$ does not strongly depend either of ${\rm NA}$ or of the core diameter (for $\omega a/c\gg 1$) and we obtained that the product $v\epsilon$ may be considered as close to $4.7$ on average, both for a \SI{50}{\micro \meter} or a \SI{105}{\micro \meter} core diameter (see~\Fig{SM_simu}).

\subsection{Expression of $\alpha$}

Along the axis defined by $\xi_+ =0$, the correlation function Eq.~\eqref{eq:correlation_eigenbasis_SM} is:
\begin{equation}
    C(0, \xi_-) \simeq \exp\left\{- \left[\frac{c}{2n_1(\omega a)^2}\right]^2 s_- \xi_-^2 \right\}
\end{equation}
Because of successive approximations carried out in the expression of eigen-vectors and eigen-values, it is simpler to express the correlation coefficient starting from Eq.~\eqref{eq:variance_ABC_SM} rather than from the expression of $s_-$. The correlation function along the axis $\delta \xi_+ =0$
may then be re-written either as a function of $\delta \omega$ or $\delta z$ in a simplified form by plugging Eq.\eqref{eq:SM_Xaxial_a} into Eq.~\eqref{eq:variance_ABC_SM}:
\begin{align}
C(0, \xi_-) &\simeq \exp\left[-\left(\frac{\delta \omega}{\Delta \omega}\right)^2\right]\\
C(0, \xi_-) &\simeq \exp\left[-\left(\frac{\delta z}{\Delta z}\right)^2\right]
\end{align}
with 
\begin{align}
    \Delta \omega &=\alpha \frac{2n_1\omega a}{{\rm NA}z}\\
    \Delta z &= \alpha(1-\epsilon) \frac{2n_1a}{{\rm NA}}.
\end{align} 
Where the expression of $\alpha$ introduced in the~\Eq{deltak} in the main text is:
\begin{equation}
    \alpha =\sqrt{\frac{v^2}{\Var(\chi u^2)-\epsilon^2 \Var(u^2)}}.
\end{equation}
We introduced $v$ together with the numerical aperture $\rm NA$ ($v=ka{\rm NA}$) in the definition of $\Delta\omega$ and $\Delta z$ in order that $\alpha$ be a dimension-less parameter.
Numerical simulations show that $\alpha$ fluctuates between $0.45$ and $1.4$ as a function of the core diameter, because of abrupt changes in modal cut-offs (as for the product $v\epsilon$).
However, assuming modal coupling in the fiber, we may consider that $\alpha$ is equal to unity on average, both for a \SI{50}{\micro \meter} or a \SI{105}{\micro \meter} core diameter as presented in~\Fig{SM_simu}.

\section{Simulation of the LP modes}

For the simulation of the LP modes we used some functions of the Matlab library of Michael Hughes~\cite{CodesMatlab}.

\medskip

\begin{figure}[t]
    \centering
    \includegraphics[width=0.9\columnwidth]{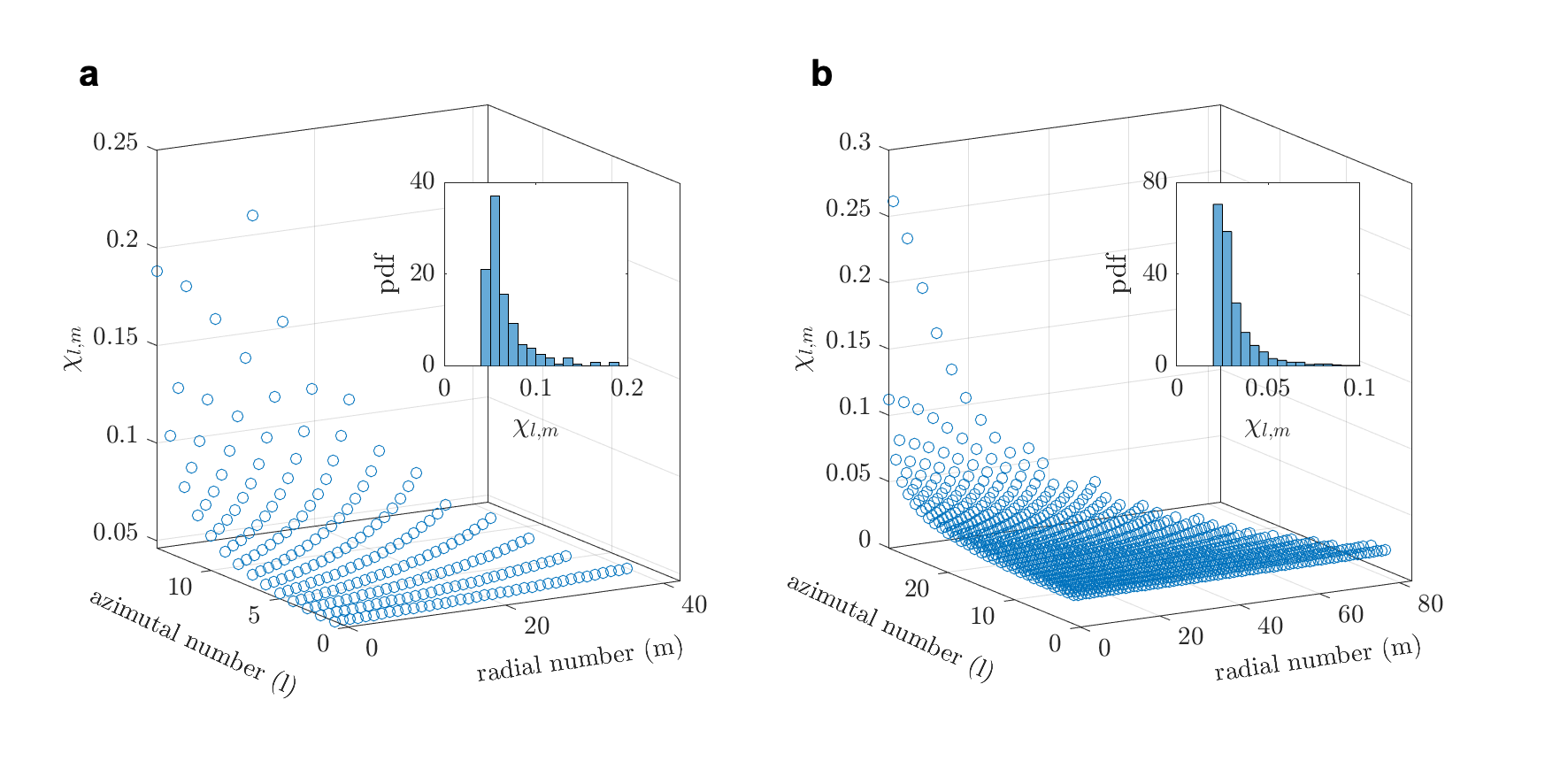}
    \caption{Distribution of $\chi_{l,m}$ to the $\chi$-axial ME for $0.22$-NA SI-MMF of diameter \SI{50}{\micro \meter} (a) and \SI{105}{\micro \meter} (b) at \SI{800}{\nano \meter}. 
    The probability density functions (pdf) presented in inset show the non-zero peaked value of $\chi$.}
    \label{fig:SM_chi}
\end{figure}

For different fibers (Ø = \SI{50}{\micro \meter} or \SI{105}{\micro \meter}), 0.22-NA, $n_{\mathrm{core}} = n_1 = 1.4533$ the modes of the fiber are numerically calculated.
The prefactors $\alpha$ and $v \epsilon$ are determined using respectively the definition linked to~\Eq{deltak} of the main text and~\Eq{SM_epsilon}.
The value of $\chi$ can be obtained by using the following equality: 
\begin{equation}
\chi_{l,m}=2\omega \frac{\partial  }{\partial \omega}\left[\ln \left( u_{l,m}\right)\right] = \frac{2}{\sqrt{v^2-(u_{l-1,m})^2 + l^2+1 }},
\label{eq:dudk}
\end{equation}
wherein for large enough core diameters ($ka\gg 1$), $u_{l,m}$ can be replaced by the $m^{th}$ zero of the Bessel function of order $l$, $J_l$~\cite{gloge1971weakly}.
The distribution of $\chi_{l,m}$ given by~\Eq{dudk} is presented in \Fig{SM_chi}a for the \SI{50}{\micro \meter} diameter fiber and \Fig{SM_chi}b for the \SI{105}{\micro \meter} diameter one. 
It appears that $\chi_{l,m}$ is almost uniform for the most confined modes (low $l$ and $m$ values) but diverges nearby the critical angle.

%%FIGURE SM?%%%%%%%%%%%%%%%%%%%%%%%%%%%%%%%%%%%%%%%%%%%%%%%%%%%%%%%
%%%%%%%%%%%%%%%%%%%%%%%%%%%%%%%%%%%%%%%%%%%%%%%%%%%%%%%%%%%%%%%%%

\begin{figure}[t]
\centering
    \includegraphics[width= 0.5\columnwidth]{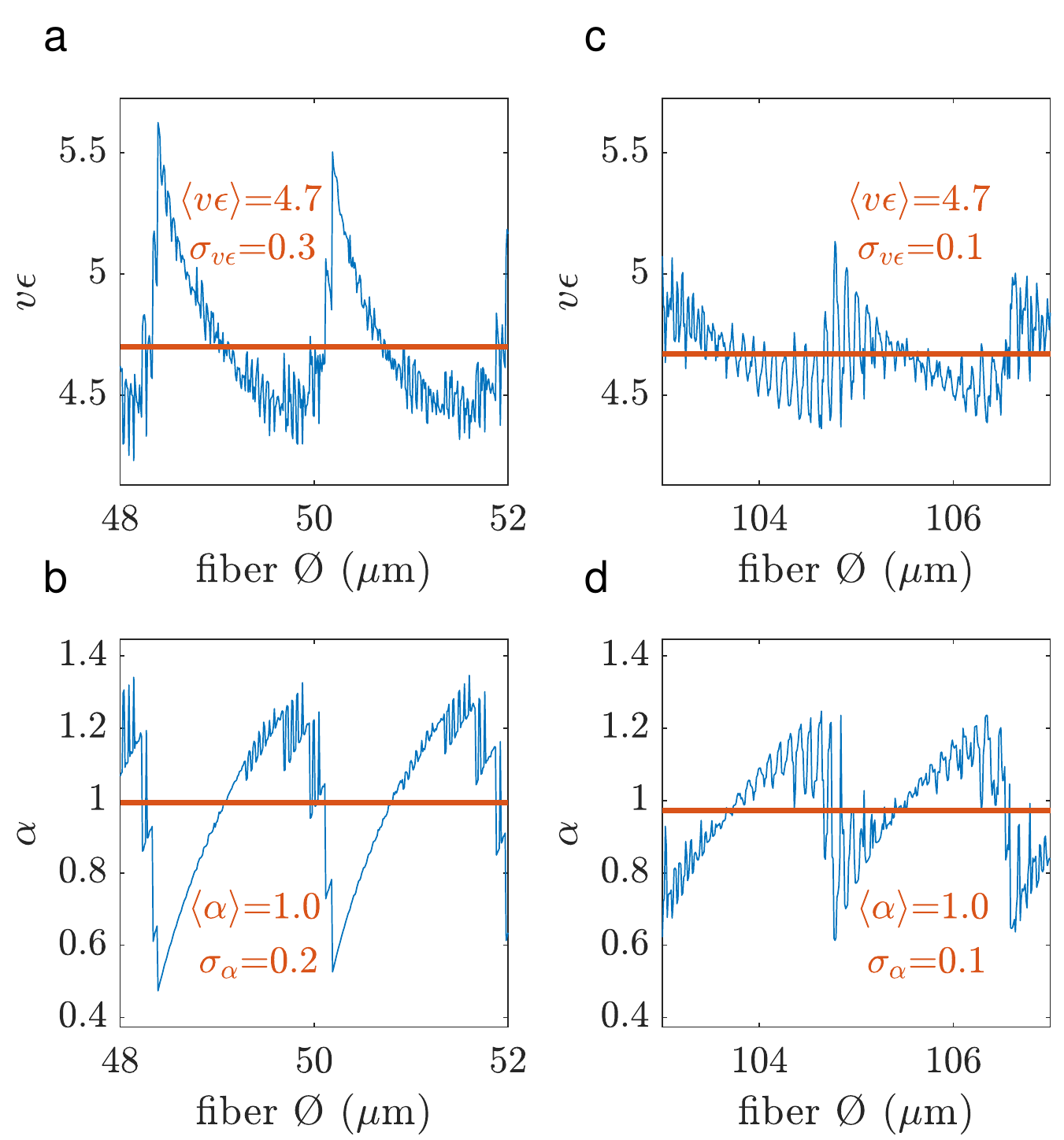}
    \caption{
    Evolution of pre-factors $v \epsilon$ (a,c) and $\alpha$ (b,d) as a function of the core diameter for 0.22-NA SI-MMF at $\lambda$ = \SI{800}{\nano \meter}. Diameters around \SI{50}{\micro \meter} (a,b) and \SI{105}{\micro \meter} (c,d) are studied.
    The mean values of $v \epsilon$ and $\alpha$ are used for the numerical simulations of the main text.
    Their standard deviation are also computed and displayed for further information.
    }
    \label{fig:SM_simu}
\end{figure}

The values of $\alpha$ and $v \epsilon$ kept for the calculations are the mean values obtained from small variations of the fiber core (due to fabrication imperfections), see~\fig{SM_simu}.

Fibers modes are then excited by an incident field.
The field is propagated along the fiber on a distance $L$ with the propagation constants of the modes previously determined.
Performing these steps for different wavelengths and fiber lengths provides the data necessary to mimic the experience.

%%FIGURE SM?%%%%%%%%%%%%%%%%%%%%%%%%%%%%%%%%%%%%%%%%%%%%%%%%%%%%%%%

%%%%%%%%%%%%%%%%%%%%%%%%%%%%%%%%%%%%%%%%%%%%%%%%%%%%%%%%%%%%%%%%%

\section{Discussion of the role of the less confined modes}

For the calculation of the phase variance in~\Eq{phase_variance} of the main text, differentiation of the product $\beta_{l,m} z$ for a given mode yields:
\begin{equation}
\delta \left(\beta_{l,m}z\right)= \delta \left( n_1\frac{\omega}{c}z\right)-\frac{u_{l,m}^2c}{2n_1 \omega^2 a^2}\left[ \omega\delta z-(1-\chi_{l,m})z\delta \omega \right]
\end{equation}
wherein it appears that the specific phase for the mode $(l,m)$, the second term, is unchanged for a $\chi$-axial shift:
\begin{equation}
    \delta z=(1-\chi_{l,m})z\frac{\delta \omega}{\omega}
\end{equation}
which slightly differs from~\Eq{Xaxial} in that $\epsilon$ is now replaced by $\chi_{l,m}$.
When calculating the phase variance driving the limited correlation range of the $\chi$-axial ME, a fine analysis of $\chi_{l,m}$ appears very informative.
As presented in~\Fig{SM_simu}, the $\chi_{l,m}$ values are almost the same for all modes except for the less confined ones.
If exciting only the most confined modes, we then get $\epsilon \simeq \chi$, with $\chi$ the common value of these confined modes, and $\alpha\rightarrow +\infty$.
As a result, if not injecting the less confined modes, responsible for field penetration in the fiber cladding, the $\chi$-axial ME extends over an infinite range in full agreement with the ray optics approximation.
Interestingly, it therefore means that the limited axial range of the $\chi$-axial ME is dominated by the less confined modes.

%\bibliography{Paper_SI_chromato_axial}

\end{document}